\documentclass[%
reprint,
superscriptaddress,
preprintnumbers,
 amsmath,amssymb,
 aps,
pre,
]{revtex4-2}

\usepackage{graphicx}% Include figure files
\usepackage{dcolumn}% Align table columns on decimal point
\usepackage{bm}% bold math
\newcommand{\MethodName}{Composite collision framework}
\newcommand{\methodName}{composite collision framework}
\newcommand{\MN}{CCF}

\newcommand{\cs}{c_{\mathrm{s}}}

\newcommand{\Iso}[1]{\delta_{#1}}
\newcommand{\wall}{\mathrm{w}}

\newcommand{\weights}{{w}}
\newcommand{\wi}{{\weights_i}}
\newcommand{\ibar}{\bar{i}}

\newcommand{\colInd}{\mathit{n}}
\newcommand{\colIndm}{\mathit{n}}

\newcommand{\etan}{\eta^\colInd}

\newcommand{\collisionSym}{\Omega}
\newcommand{\collisionBrack}[1]{\left[ #1 \right]}
\newcommand{\collisionName}[1]{ \collisionSym^\mathrm{#1}}
\newcommand{\collisionOp}[2]{ \collisionName{#1}\collisionBrack{#2} }

\newcommand{\boltzVel}{\xi}
\newcommand{\bxi}{\boldsymbol{\boltzVel}}

\newcommand{\xia}{\boltzVel_\alpha}

\newcommand{\Dt}{\Delta t}

\newcommand{\bx}{\boldsymbol{x}}
\newcommand{\xa}{x_\alpha}

\newcommand{\ff}{f}
\newcommand{\ffi}{\ff_{i}}

\newcommand{\ffieq}{{\ffi^\mathrm{eq}}}

\newcommand{\fg}{f}
\newcommand{\fgi}{\fg_{i}}
\newcommand{\fgibar}{\fg_{\ibar}}

\newcommand{\fgieq}{{\fgi^\mathrm{eq}}}
\newcommand{\fgibareq}{{\fgibar^\mathrm{eq}}}

\newcommand{\spa}{\alpha}
\newcommand{\spb}{\beta}

\newcommand{\cia}{c_{i\spa}}
\newcommand{\cib}{c_{i\spb}}

\newcommand{\bci}{\mathbf{c}_i}
\newcommand{\bcibar}{\mathbf{c}_{\bar{i}}}

\newcommand{\bvel}{\mathbf{u}}
\newcommand{\vela}{u_\alpha}
\newcommand{\velb}{u_\beta}
\newcommand{\velwa}{u_{\wall\alpha}}
\newcommand{\velwb}{u_{\wall\beta}}

\newcommand{\bflux}{\mathbf{q}}

\newcommand{\fluxwa}{q_{\wall\alpha}}

\newcommand{\force}{K}
\newcommand{\bF}{\boldsymbol{\force}}
\newcommand{\bK}{\bF}
\newcommand{\Fa}{\force_\alpha}

\newcommand{\fitA}{A}

\newcommand{\prbcFit}{\fitA}

\newcommand{\cdotnConst}{N}

\usepackage{amsmath}
\usepackage{amssymb}
\usepackage{amsfonts}
\usepackage{graphicx}% Include figure files
\usepackage{dcolumn}% Align table columns on decimal point
\usepackage{bm}% bold math
\usepackage[utf8]{inputenc}
\usepackage[T1]{fontenc}
\usepackage{mathptmx}
\usepackage{etoolbox}

\usepackage{threeparttable}

\usepackage{hyperref}% add hypertext capabilities
\usepackage[nameinlink,capitalize]{cleveref}

\usepackage{microtype}

\begin{document}

\preprint{\textit{Preprint}}

\title{Decomposing the collision operator in the lattice Boltzmann method}

\author{Julius Weinmiller}
\email{Corresponding author: julius.weinmiller@dlr.de}
\affiliation{Institute of Engineering Thermodynamics, German Aerospace Center (DLR),  Ulm, Germany}
\affiliation{Helmholtz Institute Ulm for Electrochemical Energy Storage (HIU),  Ulm, Germany}
\author{Benjamin Kellers}
\affiliation{Institute of Engineering Thermodynamics, German Aerospace Center (DLR),  Ulm, Germany}
\affiliation{Helmholtz Institute Ulm for Electrochemical Energy Storage (HIU),  Ulm, Germany}
\author{Martin P. Lautenschlaeger}
\affiliation{ SDU Mechatronics, Department of Mechanical and Electrical Engineering, University of Southern Denmark (SDU), Sønderborg, Denmark}
\author{Arnulf Latz}
\affiliation{Institute of Engineering Thermodynamics, German Aerospace Center (DLR),  Ulm, Germany}
\affiliation{Helmholtz Institute Ulm for Electrochemical Energy Storage (HIU),  Ulm, Germany}
\affiliation{Institute of Electrochemistry, Ulm University,  Ulm, Germany}
\author{Timo Danner}
\affiliation{Institute of Engineering Thermodynamics, German Aerospace Center (DLR),  Ulm, Germany}
\affiliation{Helmholtz Institute Ulm for Electrochemical Energy Storage (HIU),  Ulm, Germany}

\date{\today}% It is always \today, today,

\begin{abstract}
In transport theory, physical phenomena are well described using the Boltzmann equation, which is efficiently simulated and discretized with the lattice Boltzmann method. The collision step defines the microscopic molecules behavior, and thus the simulated physical phenomena. For complex phenomena, the collision step becomes complex as well.
In this paper, we propose a framework to systematically decompose the collision step into individual collision rules. 
Each collision rule is easier to understand, thus a faster understanding of the whole is achieved.
By inverting the process, i.e. composing multiple collision rules together, one can create novel collision steps, which can better describe the underlying complex phenomena.   
This framework's applications are manyfold, from both a theoretical and an application standpoint. Shown here is the decomposition of Robin boundary condition into Dirichlet and Neumann boundary conditions, extending it to a partial Robin boundary condition, and semi-permeable reactive membranes. 
\end{abstract}

\maketitle

\section{\label{sec:introduction}Introduction}

Complex physical phenomena are a combination of multiple processes that collectively create intricate behavior. The effects of these processes are often convoluted and cannot be easily separated. However, computational approaches are popular to study such phenomena. A prerequisite for accurate simulations is that the model's abstraction and numerical scheme captures this complexity.

The lattice Boltzmann method (LBM) is a powerful computational technique for simulating transport phenomena and related processes, known for its inherent simplicity and versatility \cite{benziLatticeBoltzmannEquation1992,krugerLatticeBoltzmannMethod2017}. It represents microscopic particles using a discrete-velocity distribution function -- the so-called population -- in a discretized phase space, using lattices, for velocity and space, respectively. The macroscopic properties and behavior emerge through the collision and streaming of populations. In LBM, simple and local collision steps can already capture fluid flow in complex geometries very well, which is one aspect making the method so popular. However, only a single collision step can be applied to a grid location. This collision step needs to describe the underlying processes. Using only simple collisions may limit the applicability of LBM, while complex collisions are difficult to comprehend and implement into simulations.

Therefore, an easier development of more complicated collision steps for complex phenomena is key to mitigate these limitations. A promising strategy is to combine simple and well-known collision rules, to create advanced and interpretable collision steps, which are able to describe complex physical phenomena. In our framework we refer to these schemes as composite collisions.

In literature, composite collisions, although not naming it as such, have been described for several phenomena. Approaches often have in common that relevant length and timescales of constitutive processes are not resolved. Most prominently, this approach has been used in the context of unresolved solid-fluid interaction, often called gray LBM or partial bounceback methods.\cite{dardisLatticeBoltzmannScheme1998,nobleLatticeBoltzmannMethodPartially1998,thorneLatticeBoltzmannModel2004,yoshidaTransmissionReflectionCoefficient2014,zhuImprovedGrayLattice2013,zhuExtendingGrayLattice2018,walshNewPartialbouncebackLatticeBoltzmann2009,gaoMesoscaleSimulationLi2023}
used to simulate single- and multi-phase flow within unresolved porous media \cite{pereiraGrayscaleLatticeBoltzmann2016, lautenschlaegerHomogenizedLatticeBoltzmann2022,lautenschlaegerUnderstandingElectrolyteFilling2022, wangImprovedPartiallySaturated2023, vienneSimulationViscousFingering2019, yuMassconservedVolumetricLattice2014, tianLatticeBoltzmannSimulation2017,sunMorphologicalHydrodynamicProperties2023,eiblLBMTwophaseBio2020}. Other applications are material dissolution \cite{petkantchinSimplifiedMesoscale3D2023,gaedtkeTotalEnthalpybasedLattice2020,liComputationalSimulationsEffects2022, anLatticeBoltzmannSimulationDissolution2020}, and (sharp) interface handling for both fluid-solid and solid-solid \cite{hanModellingThermalContact2008,liInsertedLayerLBM2021,nobleLatticeBoltzmannMethodPartially1998,minkComprehensiveComputationalModel2022,ahoDiffusionThinMembranes2016,fangNumericalPredictionsThermal2017,xieLatticeBoltzmannModeling2015}. 
%LBM literature also contains examples which are derived and treated as a single collision rules e.g.\ reactive boundary conditions \cite{Verhaeghe2006,Patel2016,Ju2020,weinmillerGeneralLocalReactive2024}.

In our previous work \cite{weinmillerGeneralLocalReactive2024} on reactive boundary conditions \cite{Verhaeghe2006,Patel2016,Ju2020,weinmillerGeneralLocalReactive2024}, we could demonstrate how careful analysis of existing boundary schemes can shed new insight into their behavior. 
Therefore, while this is not a new concept, a framework is missing for the analysis of existing, and the development of new collision steps. This is an essential step to relate complex macroscopic behavior to microscopic collisions.

In the following, we present a novel and versatile \methodName\ (\MN) for LBM. This framework provides a systematic approach to develop collision steps for complex physical phenomena by combining various simple collision rules. 
We first present the \MN\ in \cref{sec:cdf}. Collision rules used in this paper are described in \cref{subsec:collisionRules}, and continues to a discussion on forces \cref{subsec:forcing}. Finally in \cref{sec:Applications}, we apply the framework to our use cases; reinterpreting the reactive boundary condition, and presenting two new collision steps for partial reactive boundary conditions and reactive membranes.

\section{\label{sec:method} Methodology}

The LBM solves the discrete form of the continuous Boltzmann equation 
\begin{equation}
    \label{eq:boltzmannEquation}
    \frac{\partial \tilde{\ff}}{\partial t}+\xia\frac{\partial \tilde{\ff}}{\partial \xa} + \frac{\Fa}{\rho}\frac{\partial \tilde{\ff}}{\partial \xia} =\collisionSym\collisionBrack{\tilde{\ff}} +Q.
\end{equation}
The $\tilde{\ff}$ is the density distribution function, $\collisionSym$ is the collision operator, $\bF$ is the applied force density, $Q$ is a mass source, and $\bxi$ is the continuous microscopic velocity. The zeroth and first moments in $\bxi$ of $\tilde{\ff}$ are the density $\rho$ and density momentum $\rho\bvel$. Additionally, $\bx$ and $t$ denoting space and time. When using index notation, the Greek subscripts indicate space. 

\subsection{\label{sec:cdf} \MethodName}

The \MN\ is based on the idea of decomposition of the collision step. We can express any collision step as a sum of component collision rules $\collisionName{\colIndm}$, each weighted with a component fraction $\etan$
\begin{align}
    \label{eq:decomosedCollision}
    \collisionSym\collisionBrack{\tilde{\ff}} &= \sum_\colIndm \etan\collisionOp{\colIndm}{\tilde{\ff}}, & \sum_\colInd \etan =1. &
\end{align}
The aim is to choose component collision rules $\collisionName{\colIndm}$ which are simpler and better understood. This improves interpretation of $\collisionSym$ and the total behavior can be understood by the sum of its parts.

% \begin{equation}
%     \label{eq:microCD}
%     \frac{\partial \ff}{\partial t}+\frac{\partial \ff}{\partial \xa} = \sum_\colInd^N \left( \etan \collisionOp{\colInd}{\ff} - \frac{\Fa^\colInd}{\rho}\frac{\partial \ff}{\partial \xia} \right)
% \end{equation}

In LBM, the Boltzmann equation with decomposed collision step \cref{eq:boltzmannEquation,eq:decomosedCollision} is discretized to 
\begin{align}
\label{eq:MGF_main_equation}
\begin{split}
    &\ffi\left( \bx+\bci \Dt, t + \Dt \right) - \ffi(\bx, t) \\ &\qquad= \sum_{\colInd} \left( \etan \collisionOp{\colIndm}{\ffi(\bx, t)} + S_i^{\colInd}(\bx, t) \right) \Dt
\end{split}
\end{align}
Here, $\Delta t$ is the time step. The roman subscripts, e.g.\ $i$, is used as the index for velocity space. As is usual in LBM, $\ffi$ are a redefined version of $\tilde{\ff}$ which ensures second-order accurate time discretization. Then $\bci$ is the discrete microscopic velocity along the $i^{\mathrm{th}}$ lattice direction. The $S_i^\colInd$ are the mass and momentum source terms emerging from  $Q$ and $\bK$. The shorthand notation $\ffi = \ffi\left( \bx, t \right)$ is used throughout this paper.

When $\collisionOp{\colIndm}{\ffi}$ is linear in $\ffi$, as are most collision steps, then $\etan$ can be pulled into the collision rule to form
\begin{align}
    \etan\collisionOp{\colIndm}{\ffi}&\Longleftrightarrow \collisionOp{\colIndm}{\ffi^\colInd}, &
    \ffi^\colInd&:=\etan\ffi.
\end{align}
The total population is the sum of the component population $\ffi=\sum_{\colInd}{\ffi^{\colInd}}$.

This notation suggests the interpretation of $\ffi^\colInd$ as component populations. Each individual component population behavior and moments can be analyzed separately. Below are the component density $ \rho^{\colInd}$ and momentum $(\rho\bvel)^{\colInd}$, including the ``half-source'' correction emerging from the second-order time discretization.

\begin{align}
\label{eq:split_definitions_rho}
    \rho^{\colInd} :=& \sum_i f_{i}^{\colInd} + \frac{\Delta t}{2} \sum_{i} \left( 
    \collisionOp{\colIndm}{\ffi^\colInd} + S_i^{\colInd}\right)=\sum_i f_{i}^{\colInd} + \frac{1}{2}\Delta \rho^\colInd, \\
\label{eq:split_definitions_momentum}
    (\rho\bvel)^{\colInd} :=& \sum_i f_{i}^{\colInd} \bci + \frac{\Delta t}{2} \sum_{i} \left( \Omega^{\colIndm}\left[\ffi^\colInd\right] + S_i^{\colInd}\right) \bci \\ \nonumber =& \sum_i f_{i}^{\colInd} \bci + \frac{1}{2}\Delta(\rho\bvel)^\colInd,
\end{align}

The density, $\rho$, and momentum, $\rho \bvel$, of the full Boltzmann equation (cf. \cref{eq:boltzmannEquation}) are the sum of the $\colInd$-specific components via
\begin{align}
    \label{eq:total_rho_u}
    \rho &= \sum_{\colInd} \rho^{\colInd}=\sum_i \ffi + \sum_\colInd \Delta\rho^\colInd
    , \\
    \rho \bvel &= \sum_{\colInd} (\rho\bvel)^{\colInd} =\sum_i \ffi\bci + \sum_\colInd \Delta(\rho\bvel)^\colInd.
\end{align}

So far, neither $\Omega^{\colIndm}$ nor $S_i^{\colInd}$ have been defined to keep the formulation as general as possible. For example, with a single collision rule $\colInd=\mathrm{BGK}$ and $\eta^\mathrm{BGK}=1$, one recovers the classical LBGK method.
%
%This concludes the presentation of the \MN.
% The paper continues with presenting the collision rules used, a general discussion on forcing schemes for BGK, and finally applications of this framework.

%For the general case, there are several notes that should be taken into consideration.

\subsection{Common collision rules}
\label{subsec:collisionRules}
In this section, the in literature commonly seen collision rules relevant for the applications presented in this paper are reiterated. This includes: BGK collision operator with the weakly compressible equilibrium distribution, fullway bounceback (BB), equilibrium scheme (ES), anti-bounceback (ABB), and Robin boundary condition (RBC).
%Timo: In this section, some common collision rules in literature with relevance for the applications presented in the final Section are reiterated. The discussion includes: BGK collision operator with the weakly compressible equilibrium distribution, fullway bounceback (BB), equilibrium scheme (ES), anti-bounceback (ABB), and Robin boundary condition (RBC).

\paragraph{Single relaxation time (BGK)} This collision operator models transport phenomena processes as a relaxation of $\fgi$ to the equilibrium distribution $\fgieq$ with a single characteristic time $\tau$ \cite{bhatnagarModelCollisionProcesses1954}
\begin{equation}
    \label{eq:collision_BGK}
    \Omega^{\mathrm{BGK}} \left[ \fgi \right] = -\frac{1}{\tau}\left( \fgi - \fgieq\right).
\end{equation}
Fluid flow is characterized through the transport of momentum, with the equilibrium distribution defined by the moments of $\ff$. Advection of scalars, e.g.\ concentration, requires the velocities to be imposed in the equilibrium distribution.
The relaxation time $\tau$ is defined from the non-dimensional kinematic viscosity $\nu=\cs^2(\tau-\Dt/2)$ or diffusivity $D=\cs^2(\tau-\Dt/2)$, respectively, where $\cs$ is the lattice speed of sound. 

In LBM, the transport phenomena properties are defined by the equilibrium function. Different $\ffieq$ are used to characterize e.g.\ weakly compressible, linear or incompressible transport \cite{krugerLatticeBoltzmannMethod2017}. The weakly compressible equilibrium function commonly used in the collision above is given as 
\begin{align}
    \fgieq(\rho,\bvel) = \wi\rho\left( 1 + \frac{\cia \vela}{\cs^{2}} + \frac{ \vela \velb \left(\cia \cib - \cs^2 \Iso{\spa\spb}\right) }{2\cs^{4}}\right),
\label{eq:fluid_Equilib}
\end{align}
where $\lbrace \wi \rbrace$ is the lattice weight set, derived from the discretized Boltzmann distribution \cite{krugerLatticeBoltzmannMethod2017}. Here, $\Iso{\spa\spb}$ is the Kronecker delta. %The lattice weight set has to fulfill the conservation of mass, momentum and rotational isotropy \cite{krugerLatticeBoltzmannMethod2017}. 
For clarity, the density and velocity used to compute $\ffieq$ can be denoted as $\rho^\mathrm{eq}$ and $\bvel^\mathrm{eq}$.

\paragraph{Fullway bounceback (BB)} The bounceback boundary condition used to implement static no-slip or adiabatic walls, i.e. zero velocity Dirichlet boundary condition, is given as \cite{ginzbourgLocalSecondorderBoundary1996, laddNumericalSimulationsParticulate1994}
\begin{align}
    \label{eq:bouncebackScheme}
    \fgibar(\bx,t+\Dt)&=\fgi^\star(\bx,t),
\end{align}
where $\fgi^\star(\bx,t)$ is the post-collision state. The notation $\bar{i}$ indicates the opposite direction of $i$, i.e.~$\bcibar=-\bci$. Following \cite{nobleLatticeBoltzmannMethodPartially1998,krugerLatticeBoltzmannMethod2017}, the solid boundary condition is expressed as a local collision rule, such that it is compatible with \cref{eq:MGF_main_equation}. The bounceback boundary condition can be approximated as a collision rule within the wall
\begin{equation}
    \label{eq:collision_BB_velocity}
    \collisionOp{BB}{\fgi}\Dt =  -\fgi + \fgibar + 2\wi \rho_\wall \frac{\cia \velwa}{\cs^2},
\end{equation}
moving with velocity $\bvel_\wall$, where the subscript $\wall$ is referring to the properties at the wall.

This is known as the fullway bounceback version and is simpler to implement in code. The unknown $\fgi$ are determined in the wall next to the interface, and then streamed into the fluid domain in the next timestep \cite{krugerLatticeBoltzmannMethod2017}. This, however, introduces an time delay of one $\Dt$. 
%The bounceback scheme (cf.\ \cref{eq:bouncebackScheme}) itself is usually implemented during the streaming, thus within the same timestep, and better for transient processes \cite{krugerLatticeBoltzmannMethod2017}. It can also be implemented as a non-local collision rule \cite{thorneLatticeBoltzmannModel2004,yoshidaTransmissionReflectionCoefficient2014}.

With the same scheme, it is possible to approximate a mass or concentration density Neumann boundary condition to a velocity Dirichlet boundary condition for advection-diffusion processes \cite{ginzburgGenericBoundaryConditions2005,krugerLatticeBoltzmannMethod2017}. The resulting approximated collision rule is 
\begin{equation}
    \label{eq:collision_BB_flux}
    \collisionOp{BB}{\fgi} \Dt=  -\fgi + \fgibar + 2 \wi \frac{\cia \fluxwa}{\cs^2},
\end{equation}
where $\bflux_\wall$ is the applied flux of the Neumann boundary condition.

\paragraph{Fullway anti-bounceback (ABB)} The anti-bounceback boundary condition \cite{ginzburgGenericBoundaryConditions2005,TRTGinzburg2008, TRTGinzburg2008a, izquierdoCharacteristicNonreflectingBoundary2008} is a popular method to describe a density Dirichlet boundary condition in LBM. Similar to the fullway bounceback, it can be expressed as a collision rule in the adjacent cell
\begin{equation}
\begin{split}
    \label{eq:collision_ABB}
  &\collisionOp{ABB}{\fgi}\Dt = -\fgi - \fgibar\\& \qquad+ 2\wi \rho_\wall \left( 1 + \frac{ \velwa \velwb \left( \cia \cib - \cs^2 \Iso{\spa\spb} \right) }{2\cs^{4}} \right).
\end{split}
\end{equation}

\paragraph{Equilibrium scheme (ES)} This simple scheme \cite{lattStraightVelocityBoundaries2008,heAnalyticSolutionsSimple1997} is a method of describing both density and velocity by defining the populations directly by its equilibrium. Hence, its collision rule is given by
\begin{equation}
    \label{eq:collision_equilib}
    \collisionOp{ES}{\fgi}\Dt = -\fgi + \fgieq. 
\end{equation}
%While the method is very stable, it is only accurate to first order \cite{krugerLatticeBoltzmannMethod2017}.
Note that the $\rho^\mathrm{eq}$ and $\bvel^\mathrm{eq}$ are imposed, and not necessarily that of the fluid.
This collision rule can also be written in terms of the $\collisionName{ABB}$ \cref{eq:collision_ABB} and $\collisionName{BB}$ \cref{eq:collision_BB_velocity}. 
Expressing $\fgieq$ (cf.\ \cref{eq:fluid_Equilib}) in terms of symmetric (+) and antisymmetric parts (-) in velocity space \cite{ginzburgGenericBoundaryConditions2005} gives
\begin{align}
    \label{eq:feqsym}
    \fgieq &= \fgieq^+ + \fgieq^-, & \text{with} &  & \fgieq^\pm&=\frac{\fgieq \pm \fgibareq}{2}.
\end{align}
They have the properties that 
\begin{align}
    \label{eq:symProp}
     \fgibareq^+ &= \fgieq^+, && \text{and}& \fgibareq^- &= -\fgieq^-.
\end{align}
In this specific case, $\fgieq^+$ contains the even-order velocity terms and $\fgieq^-$ the odd-order velocity terms of $\fgieq$ (cf.\ \cref{eq:fluid_Equilib}).
Then $\collisionName{ABB}$ and $\collisionName{BB}$ are given by
\begin{align}
    \collisionOp{ BB}{\fgi} \Dt&= -\fgi + \fgibar + 2\fgieq^-,\\
    \collisionOp{ABB}{\fgi} \Dt&= -\fgi - \fgibar + 2\fgieq^+.
\end{align}
From this follows that 
\begin{equation}
    \label{eq:ES_decomposed}
    \frac{\collisionOp{ABB}{\fgi} + \collisionOp{ BB}{\fgi}}{2} = \collisionOp{ES}{\fgi}.
\end{equation}
This relation shows that the equilibrium scheme $\collisionName{ES}$ can be decomposed into $\collisionName{ABB}$ and $\collisionName{BB}$ with equal proportions -- it is both a density and velocity Dirichlet boundary.
Additionally, subtracting $\collisionName{BB}$ from $\collisionName{ABB}$, and using \cref{eq:symProp}, results in the following relation
\begin{equation}
    \label{eq:ABB_minus_BB}
    \frac{\collisionOp{ABB}{\fgi} - \collisionOp{ BB}{\fgi}}{2} = \collisionOp{ES}{\fgibar}.
\end{equation}

\paragraph{Robin boundary condition (RBC)} This boundary condition is a combination of both a density Dirichlet and Neumann boundary condition. 
It describes the mass flux for density fields, first-order reactions for concentration fields, or convection boundaries for temperature fields, with the relevant macroscopic equation
\begin{equation}
\label{eq:flux_macro}
    \bflux_\wall = -D\frac{\partial\rho}{\partial \mathbf{x}} = k_\mathrm{r} \left(\rho_\wall^\mathrm{eq} - \rho  \right).
\end{equation}
Here, $k_\mathrm{r}$ is the transfer coefficient and $\rho_\wall^\mathrm{eq}$ is the equilibrium at the wall. 
It models behavior between zero flux, $k_\mathrm{r}=0$, and infinitely fast, $k_\mathrm{r}\rightarrow \infty$, transfer rate \cite{Verhaeghe2006,Patel2016,Ju2020,weinmillerGeneralLocalReactive2024,huangBoundaryConditionsLattice2015}. 

For a general local first-order equilibrium reaction without velocity \cite{Verhaeghe2006,Patel2016,Ju2020}, it was shown that a unified formulation can be derived \cite{weinmillerGeneralLocalReactive2024}. Here, we show the scheme found in literature, and its rewritten form as a collision rule
\begin{flalign}
    \fgi^\star &= \frac{2k_i}{1+k_i} \wi \rho_\wall^\mathrm{eq} + \frac{1-k_i}{1+k_i} \fgibar, \\
    \label{eq:collision_RBC}
    \collisionOp{RBC}{\fgi} \Dt &= -\fgi + \frac{2k_i}{1+k_i} \wi \rho_\wall^{\mathrm{eq}} + \frac{1-k_i}{1+k_i} \fgibar.
    %\label{eq:collision_RBC}
    %\Omega^\mathrm{RBC} \left[ \ffi^{\colInd} \right] &= \frac{2k_i}{1+k_i}\Omega^\mathrm{ES} \left[ \ffi^{\colInd} \right] + \frac{1-k_i}{1+k_i} \Omega^\mathrm{BB} \left[ \ffi^{\colInd} \right].
\end{flalign}
The $k_i = \gamma\, k_{\mathrm{r}}\,(\bcibar\cdot\mathbf{n}) / \cs^{2} $ is the directional transfer rate, $\mathbf{n}$ is the wall normal pointing into the fluid, and $\gamma=\tau / (\tau-\Dt/2)$ is a diffusion correction term. The different schemes in literature differ slightly in the definition of $k_i$.

\subsection{Forcing in composite collisions}
\label{subsec:forcing}

The force density in LBM is effectively a momentum source
\begin{equation}
    \label{eq:pureForceMomentumShift}
    \Delta(\rho\bvel)^\colInd=\bK^\colInd\Dt.
\end{equation}
There are two central questions for the inclusion of forces in the \methodName: 1) how are force densities decomposed and 2) how do forcing terms manifest for the composite collision components $\colInd$.

The general answer to 1) is 
\begin{equation}
    \label{eq:proportionalForce}
    \bF^\colInd=\etan\bF.
\end{equation}
A straightforward argumentation is that one actually applies an acceleration field $\mathbf{a}$, which all components experience equally $\mathbf{a}^\colInd=\mathbf{a}$. Thus $\bF = \rho \mathbf{a}$ and hence $\bF^\mathbf{\colInd} = \rho^\mathbf{\colInd} \mathbf{a}$. 

For the second question 2), it depends on the composite collision components $\colInd$. Additionally, the forcing terms can manifest indirectly through the the equilibrium velocity ${\bvel^\mathrm{eq}}^\colInd$, or directly via $S_i^\colInd$.
Here we will discuss it first for the special case of velocity boundary conditions, at the hand of $\collisionName{BB}$, and then for the general case, taking $\collisionName{BGK}$ as a final example.

The $\collisionName{BB}$ is a velocity Dirichlet boundary condition, where the velocity is prescribed. Thus the momentum source due to the force density (cf.\ \cref{eq:pureForceMomentumShift}) is overwritten and resulting in ${\bvel^\mathrm{eq}}^\mathrm{BB}=\bvel_\wall$ and $S^\mathrm{BB}_i = 0$.
A different argumentation from microscopic perspective is that the bounceback populations ``travel'' $\Dt/2$ with the accelerating field and $\Dt/2$ against, thus canceling out the effect.

Continuing with the general case, forcing schemes in LBM can be differentiated in their source term $S_i^\colInd$ and how they modify the equilibrium velocity
\begin{equation}
    \label{eq:forcesEquilibriumVelocity}
    {\bvel^\mathrm{eq}}^\colInd = \frac{1}{\rho^\colInd} \sum_i\ffi^\colInd\bci + B \frac{\bF^\colInd\Dt}{\rho^\colInd},
\end{equation}
where $B$ is a parameter depending on the forcing scheme. 
%Most forcing schemes recover second order accuracy, and differing in higher order terms. 
We will elucidate \MN\ properties by analyzing both, starting with the ${\bvel^\mathrm{eq}}^\colInd$.

In the absence of mass sources $\Delta \rho^\colInd=0$, the relation $\rho^\colInd=\etan \rho$ holds and one can simplify \cref{eq:forcesEquilibriumVelocity} to 
\begin{equation}
    \label{eq:forcesEquilibriumVelocitySimple}
    {\bvel^\mathrm{eq}}^\colInd = \frac{1}{\rho} \sum_i\ffi\bci + B \mathbf{a}^\colInd\Dt,
\end{equation}
where $\mathbf{a}^\colInd=\bF^\colInd/\rho^\colInd$ is the acceleration experienced and $\mathbf{a}^\colInd = \mathbf{a}=\bF/\rho$. Immediately, we can see that the velocity is not dependent on the $\etan$. 

With a lack of mass sources, the term $S_i^\colInd$ is solely a momentum source due to forces and $\sum_i S_i^\colInd = 0$. Given \cref{eq:proportionalForce,eq:pureForceMomentumShift}, the source term should be proportional to the composite fraction $S_i^\colInd \propto \etan$ for $\colInd \neq \mathrm{BB}$. A simple thought experiment to show that is as follows; lets assume the collision is momentum conserving $\sum_i \collisionOp{\colInd}{\ffi^\colInd}\bci=0$, then that leads to $\sum_i S_i^\colInd\bci=\bK^\colInd=\etan\bK$. Therefore $S_i^\colInd \propto \etan$.

For a complete example, let us look at $\collisionName{BGK}$. The resulting momentum with that new equilibrium velocity (cf.\ \cref{eq:forcesEquilibriumVelocitySimple}) is
\begin{align}
\sum_i\collisionOp{BGK}{\ffi^\mathrm{BKG}}\cia
&= -\frac{1}{\tau}\sum_i\left( \ffi^\mathrm{BGK} - \ffieq^\mathrm{BGK}  \right) \cia,  \nonumber \\
&= -\frac{\eta^\mathrm{BGK}}{\tau}\left( \sum_i(\ffi\cia)  - \rho \vela^\mathrm{eq}\right),  \nonumber \\
&= \frac{\eta^\mathrm{BGK}\rho}{\tau} B {a}_\spa \Dt.
\end{align}
Due to the second-order time discretization with forces, the BGK collision operator is not momentum conserving, and we have a momentum shift \cite[ch.6]{krugerLatticeBoltzmannMethod2017}. This can be used directly, i.e. $B=\tau$ to immediately achieve the wanted force $\bF^\mathrm{BGK}=\eta^\mathrm{BGK} \rho \mathbf{a}$.

Including the source term $S_i^\mathrm{BGK}$ to get the complete momentum source, we get
\begin{align}
    \sum_i\left( \collisionOp{BGK}{\ffi^\mathrm{BKG}} + S_i^\mathrm{BGK} \right)\cia  =\Fa^\mathrm{BGK}  \nonumber\\
    \frac{\eta^\mathrm{BGK}\rho}{\tau} B {a_\spa} \Dt  + \sum_i S_i^\mathrm{BGK}\cia  =\Fa^\mathrm{BGK} \nonumber\\
    \sum_i S_i^\mathrm{BGK}\cia = \left( 1 - \frac{B\Dt}{\tau} \right)\Fa^\mathrm{BGK}
\end{align}

To conclude, forcing terms in \MN\ only need minor modifications. In general, the velocity is independent of the composite fraction, and is equal for all component parts. The only location introducing the composite fraction is at the source term. These rules do not apply to collision steps which prescribed velocities, such as BB which is a velocity Dirichlet boundary conditions.

\section{\label{sec:Applications}Application}
In this section we will present several applications and combinations of the shown methods.
First, the \methodName\ is applied to analyze a RBC \cref{subsec:RBCrewrite}, providing key insights and demonstrating a straightforward way to extend it. Afterwards, two new applications are introduced, covering partially reactive walls \cref{subsec:prbc} and reactive membranes \cref{subsec:membrane}

\subsection{Analysis of advection-diffusion flux boundary conditions}
\label{subsec:RBCrewrite}
In this section, we will show how \MN\ can be applied to rewrite existing collision steps. We show this procedure by analyzing a Robin-type boundary condition (cf.~\cref{eq:collision_RBC}). The aim is to provide on the one hand better physical insight to the RBC in LBM and on the other hand show how rewriting can be advantageous for implementation of complex operators.

\paragraph{Physical insight} The RBC is a linear combination of the Dirichlet and Neumann boundary condition. From \cref{subsec:collisionRules}, the Dirichlet boundary condition set via $\Omega^\mathrm{ABB}$ and the Neumann boundary condition can be approximated by $\Omega^\mathrm{BB}$. 

The $\Omega^\mathrm{RBC}$ (cf.\ \cref{eq:collision_RBC}) recovers in its extremes of $k_i = 0$ and $k_i\rightarrow\infty$ the $\Omega^\mathrm{BB}$ and $\Omega^\mathrm{ABB}$ collision steps, respectively. Using those as the basis for a composite collision gives
\begin{align}
    \label{eq:RBC_decomposed_GCD}
    \collisionOp{RBC}{\ffi} = \sum_{n\in C} \collisionOp{\colInd}{\ffi^\colInd} = \collisionOp{ABB}{\ffi^\mathrm{ABB}} + \collisionOp{BB}{\ffi^\mathrm{BB}},
\end{align}
with the collision set $C=\{\mathrm{BB},\ \mathrm{ABB}\}$.

From the \MN\ the following relations hold, $\eta^\mathrm{BB}+\eta^\mathrm{ABB}=1$ and $\rho_\wall^\colInd = \eta^\colInd\rho_\wall$.
Hence, due to the linearity of the collision steps, one can factor out the $\eta^\colInd$, and \cref{eq:RBC_decomposed_GCD} becomes
\begin{align}
    \label{eq:RBC_decomposed_simple}
    \collisionOp{RBC}{\ffi} &= \eta^\mathrm{ABB} \collisionOp{ABB}{\ffi} +\eta^\mathrm{BB} \collisionOp{BB}{\ffi}.
\end{align}

Then using the decomposed ES formulations (cf.~\cref{eq:ES_decomposed}),
allows for combining \cref{eq:RBC_decomposed_simple,eq:ES_decomposed} and writing the composite RBC as
\begin{align}
    \collisionOp{RBC}{\ffi} &= 2\eta^\mathrm{ABB} \collisionOp{ES}{\ffi} + (\eta^\mathrm{BB}-\eta^\mathrm{ABB}) \collisionOp{BB}{\ffi}.
    \label{eq:RBC_MN_collision}
\end{align}

To relate this composite collision formulation of the RBC with the one from literature (cf.~\cref{eq:collision_RBC}), it needs to be rewritten. Realizing that the two fractions in the scheme always add to one, i.e. ${(2k_i)}/{(1+k_i)} + {(1-k_i)}/{(1+k_i)} = 1$, and for simplicity setting $\bci\cdot\mathbf{n}=\cdotnConst$ to a constant, allows splitting the $-\fgi$ term, and expressing the literature RBC (cf.~\cref{eq:collision_RBC}) as a composite collision
\begin{equation}
    \label{eq:collision_RBC_rewritten}
    \collisionOp{RBC}{\fgi} = \sum_{n\in C} \collisionOp{\colInd}{\ffi^\colInd} =  \frac{2k_i}{1+k_i} \collisionOp{ES}{\fgi} + \frac{1-k_i}{1+k_i} \collisionOp{BB}{\fgi},
\end{equation}
with the collision set $C=\{\mathrm{ES},\ \mathrm{BB}\}$.

Assuming that the RBC is the sole collision step, allows comparing \cref{eq:collision_RBC_rewritten,eq:RBC_MN_collision} for the computation of the composite fractions in \cref{eq:RBC_decomposed_simple}. They are $\eta^\mathrm{ABB} = k_i / ( 1+ k_i )$ and $\eta^\mathrm{BB} = 1 / (1+k_i)$. Inserting the definition of $k_i$ results in 
\begin{equation}
    \label{eq:RBC_eta}
    \eta^\mathrm{BB} = \frac{1}{\tau \cdotnConst k_\mathrm{r}/ D + 1 }.
\end{equation}
A short discussion on the similarity of this bounceback composite fraction to the partial bounceback methods is given in \cref{app:compareCompositeBB}. When using the $\collisionName{RBC}$ as a fullway boundary collision, the wall normal can be reinserted $\cdotnConst=\bci\cdot\mathbf{n}$. 

Splitting the $\collisionName{RBC}$ in such a fashion allows for an easier analysis. Often the question is how much concentration/mass/heat was added to the system. This is purely driven by the $\collisionName{ABB}$ only, and the $\collisionName{BB}$ does not contribute, i.e. $\Delta\rho^\mathrm{BB}=0$ and $\Delta\rho=\Delta\rho^\mathrm{ABB}$.

\paragraph{Implementation} From a code implementation point of view, the RBC collision step (cf. \cref{eq:RBC_decomposed_simple}) requires adding a new collision function block. However, using \MN, it can be implemented in a way that reuses code. Starting at \cref{eq:RBC_decomposed_simple}, we can rewrite \cref{eq:ABB_minus_BB} as $ \collisionOp{ABB}{\fgi} = \collisionOp{BB}{\fgi} + 2 \collisionOp{ES}{\fgibar}$. Together, they form 
\begin{align}
    \label{eq:RBC_decomposed_simple_impl}
    \collisionOp{RBC}{\fgi} 
    &= \eta^\mathrm{ABB}\left( \collisionOp{BB}{\fgi} + 2 \collisionOp{ES}{\fgibar} \right) +\eta^\mathrm{BB}\collisionOp{BB}{\fgi}, \\
    \label{eq:RBC_BB_with_source}
     &= \collisionOp{BB}{\fgi} + 2 \eta^\mathrm{ABB} \collisionOp{ES}{\fgibar}.
\end{align}
In this rewritten form, we can see that the $ \collisionOp{RBC}{\fgi}$ can be expressed as a bounceback with a source. The source is dependent on $ \collisionOp{ES}{\fgibar} = \fgibareq - \fgibar $, which is analogous to the $(\rho_\mathrm{w}^\mathrm{eq} - \rho)$ in the first order flux equation (cf.\ \cref{eq:flux_macro}). It vanishes as the density difference approaches the equilibrium density.
%Additionally, several parallels can be drawn between $k_i ( \fgibareq - \fgibar ) $ and the velocity term of the bounceback $2 \wi \bci\cdot \mathbf{q_\wall} / \cs^2 $  (cf.\ \cref{eq:collision_BB_flux}). 

To conclude this part, using the \MN, the RBC was broken down into simpler, better understood collision steps. This analysis shows how
1) this RBC can be extended to include a wall velocity, by using the full BB and ABB equations \cref{eq:collision_BB_velocity,eq:collision_ABB}.
2) physically equivalent operators, such as NEBB or interpolated BB, can be used to generate new RBC collision steps.
3) the RBC can be rewritten to a more code implementation friendly formulation, which should help in the speed of development.

\subsection{Partial Robin boundary condition}
\label{subsec:prbc}
In this example, the \MN\ is used to generate a new collision step -- the partial RBC (PRBC).
This collision can be used to model sub-grid effects. The nucleation process is a good illustrative physical example. A nucleus' size is usually just a few nanometers, typically much smaller than the grid size of the simulation. Hence, a grid point is better simulated as a partially reactive wall, where the majority of the surface is non-reactive.

The PRBC effectively models a wall where only a section has some flux. This can be interpreted as an imposition of reactive and non-reactive surfaces. Reactive flows are used here for illustration purposes, but this is valid for mass or heat flows as well. The impact of this imposition is discussed near the end of this section.

The two collision rules being combined are $\collisionName{RBC}$, discussed in \cref{subsec:RBCrewrite}, and $\collisionName{BB}$. The resulting collision step reads 
\begin{equation}
    \collisionOp{PRBC}{\ffi} =\sum_{n \in C} \collisionOp{\colInd}{\ffi^\colInd} =\collisionOp{RBC}{\ffi^\mathrm{RBC}} + \collisionOp{BB}{\ffi^\mathrm{BB}},
\end{equation}
with the collision set $C=\{\mathrm{RBC},\ \mathrm{BB}\}$.

\begin{figure}[t]
    \centering
    \includegraphics[width=0.7\linewidth]{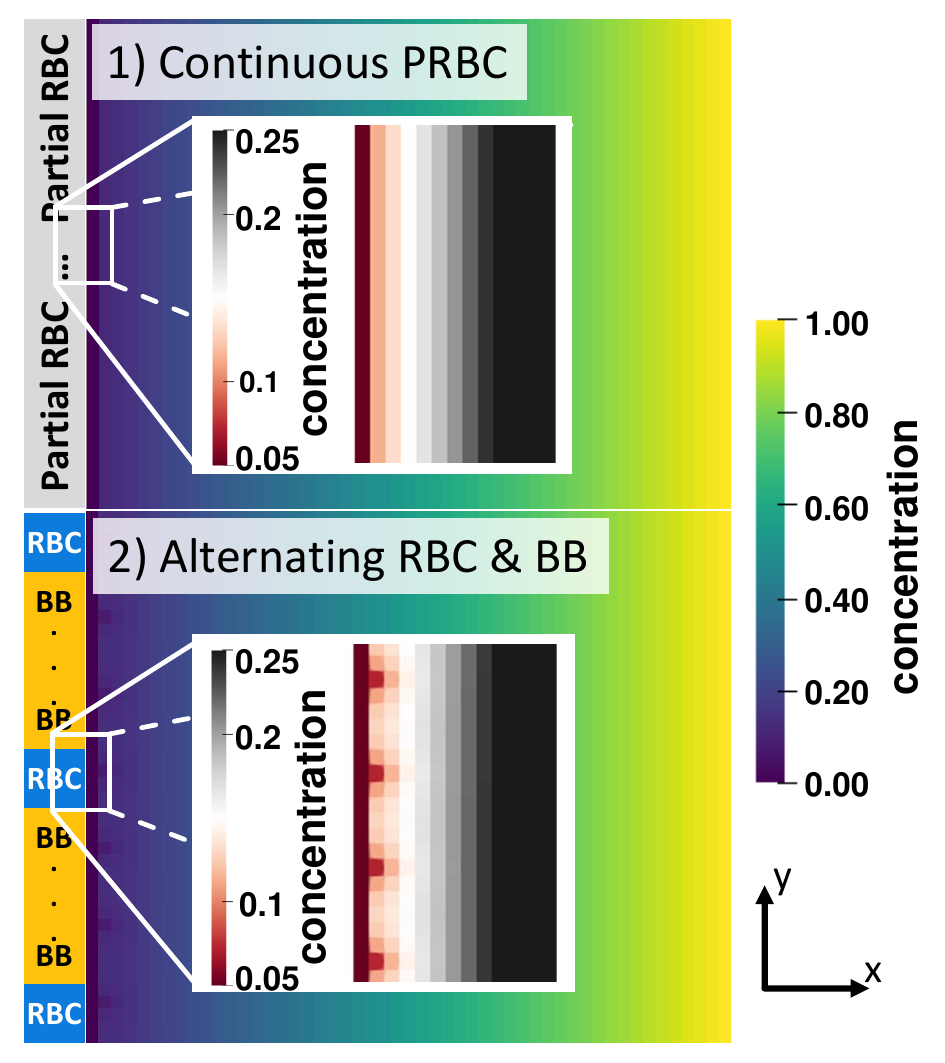}
    \caption{Steady-state simulation results comparing the novel PRBC to resolved BB-RBC setup. A spacing of $N_\mathrm{BB}=5$ with a $\mathrm{Da}=500$ is used. For the PRBC, the composite fraction used is $\eta^\mathrm{RBC}=1/9$. The inset shows the behavior near the wall to highlight the effects in that region.}
    \label{fig:PRBCsetup}
\end{figure}

In the following, a pure diffusion problem is solved ($\mathbf{u}=0$) which is used to compare a fully resolved alternating $\collisionName{RBC}$ and $\collisionName{BB}$ boundary versus the proposed composite $\collisionName{PRBC}$. A visual representation is shown for one particular case in \cref{fig:PRBCsetup}, with a domain of $50\times 200$. The Péclet number ($\mathrm{Pe}=LU/D$) is 1, and lattice velocity is $U=0.02$. The reaction is defined via the Damköhler number ($\mathrm{Da}=k_\mathrm{r}/U$) and $k_\mathrm{r}$ is varied. At the right boundary ($x=50$), a density Dirichlet condition is set with $\rho=1$, the domain in the y-axis is periodic. In our nucleation example, $\rho$ is representing concentration. The left boundary ($x=0$), either consists of 1) the proposed continuous PRBC or 2) an alternating only reactive -- only BB boundary condition.

The number of fully BB grid points between each RBC grid point is $N_\mathrm{BB}$ and will be used in the computation of $\eta^\mathrm{PRBC}$. An example simulation comparing the two methods is given in \cref{fig:PRBCsetup}, where the top half is the novel PRBC and the bottom half alternating fully reactive.

\begin{figure}[t]
    \centering
    \includegraphics[width=0.7\linewidth]{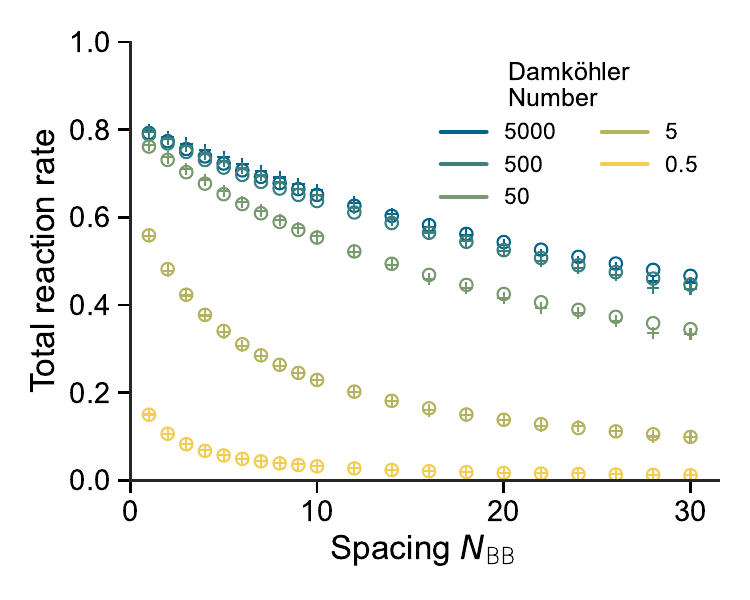}
    \caption{Comparison of non-dimensional reaction rates from boundary condition using the alternating (pluses) vs PRBC (circles). The spacing $N_\mathrm{BB}$ is varied between 1 and 30. The Damköhler numbers $\mathrm{Da}=\{0.5, 5, 50, 500, 5000\}$ were simulated, where the fitting parameter $\prbcFit=\{1, 1, 1.3, 1.9, 2.0\}$ were used, respectively.}
    \label{fig:PRBCvariations}
\end{figure}

The comparison in \cref{fig:PRBCsetup} shows a spacing of $N_\mathrm{BB}=5$ with a $\mathrm{Da}=500$ and corresponding $\eta^\mathrm{RBC}=1/9$ to match the alternating case. Other Damköhler numbers, spacing, and $\eta^\mathrm{RBC}$ values were also simulated and plotted in \cref{fig:PRBCvariations}.
For these simulations, the composite fraction is given by
\begin{equation}
    \label{eq:etaPRBC}
    \eta^\mathrm{RBC} = \frac{1}{\prbcFit N_\mathrm{BB}+1}.
\end{equation}
Here, $\prbcFit$ is a parameter to fit the composite to the resolved simulation result. Defining the composite fraction in this way is in line with approaches previously reported in the literature (cf.~\cref{eq:eta_Walsh,eq:RBC_eta}). Since only one reactive grid point is simulated $N_\mathrm{RBC}=1$, the impact of multiple reactive grid points is not shown. However logically, the relevant factor is $N_\mathrm{BB} / N_\mathrm{RBC}$. This is true especially for the reaction limited case ($\mathrm{PeDa}\ll1$), where the exact positions of the reactive grid points are negligible.

Due the wall length being a fixed 200 grid points, certain $N_\mathrm{BB}$ spacings have the same number of fully reactive grid points. For example $N_\mathrm{BB}=\{22,\ 24\}$, and $N_\mathrm{BB}=\{28,\ 30\}$ have 9 and 7 reactive grid points, respectively. However, for the higher spacing of each of the two pairs, the distribution is slightly worse, hence decreasing the diffusion towards the reactive grid points and thus slightly decreasing total reactions, as can be seen in \cref{fig:PRBCvariations}.

From \cref{fig:PRBCvariations}, it is clear that the relation of \cref{eq:etaPRBC} captures the behavior, as long as an appropriate value for $\prbcFit$ is used. A sensitivity analysis of $\prbcFit$ is given in Appendix\ \ref{app:PRBCsensitivity}. 

The parameter $\prbcFit$ is related to how reaction- or diffusion-limited the system is, with reaction limit $\prbcFit=1$ and diffusion limit $\prbcFit \rightarrow \prbcFit_\mathrm{lim}$. The $\prbcFit_\mathrm{lim}$ is a constant related to the system geometry, which in our case is approximately $2$.
With some algebraic manipulation, \cref{eq:etaPRBC} can be rewritten to
\begin{equation}
    \label{eq:etaRbcAreaFraction}
    \eta^\mathrm{RBC}= \frac{1}{ \prbcFit N_\mathrm{BB} /  N_\mathrm{RBC} +1 } =  N_\mathrm{RBC}/(\prbcFit N_\mathrm{BB}+ N_\mathrm{RBC}).
\end{equation}

When $\prbcFit=1$, \cref{eq:etaRbcAreaFraction} shows that $\eta^\mathrm{RBC}$ is the area fraction of the reactive substrate. 
The parameter $\prbcFit$ is then an area correction factor needed for the diffusion limited case.
That is a consequence of the imposition. It assumes that the time scale within the grid point is very fast, and thus negligible. That results in an effectively infinitely fast diffusion within the PRBC grid point. In case the setup is diffusion limited, the time scale becomes relevant, which can be represented as if there are effectively fewer $N_\mathrm{RBC}$ or more $N_\mathrm{BB}$. 

% In this paper, it is only shown that a relation for $\eta^\mathrm{RBC}$ exists. The exact formulation is beyond the scope of this work.

%Overall, this new PRBC can find applications where the boundary behavior is driven by a different length scale used in the bulk. Such as nuclei or nano-particles on substrates for reactions, injectors for mass fluxes in large chambers, or conductive fibers in non-conductive resin. 

\subsection{Porous-media with RBC surfaces}
\label{subsec:membrane}

This showcase presents a diffusion case through a porous media with surfaces that are described by the RBC. The porous media is modeled via the partial bounceback \cite{walshNewPartialbouncebackLatticeBoltzmann2009}.
Relevant examples are concentration diffusion through a reactive membrane or particulate mass flow through a dust filter. Macroscopically, the relevant processes are advection-diffusion (BGK) and Robin boundary condition (BB+ABB).
In the \MN\ formulation, they results in a partial bounceback with fluxes (PBBF) collision step given as
\begin{equation}
    \label{eq:collision_PBBF}
    \Omega^{\mathrm{PBBF}}\left[ f_i \right] =\sum_{n \in C} \collisionOp{\colInd}{\ffi^\colInd}  = \Omega^{\mathrm{RBC}}\left[f_i^\mathrm{RBC}\right] + \Omega^{\mathrm{BGK}}\left[f_i^\mathrm{BGK}\right],
\end{equation}
with the collision set $C=\{\mathrm{RBC},\ \mathrm{BGK}\}$.

In the limit of no flux, $k_\mathrm{r}=0$ (cf.~\cref{eq:RBC_eta}), the RBC recovers the BB collision rule, thus the PBBF results in the known partial bounceback BGK-BB method \cite{walshNewPartialbouncebackLatticeBoltzmann2009}. For all transfer coefficients $k_\mathrm{r}>0$, it is a new collision step.

Intuitively, we can assume that a mixture of flow, wall and fluxes should result in some kind of semi-permeable PBB with sources. 
Using the implementation friendly formulation of RBC (cf.\ \cref{eq:RBC_BB_with_source}), we can indeed rewrite PBBF into
\begin{equation}
    \label{eq:collision_PBBF_source}
    \Omega^{\mathrm{PBBF}}\left[ f_i \right] = \Omega^\mathrm{PBB}\left[ f_i \right] + 2\eta^\mathrm{RBC} \eta^\mathrm{ABB} \Omega^\mathrm{ES}\left[ f_{\bar{i}} \right].
\end{equation}

\begin{figure}[t]
    \centering
    \includegraphics[width=0.7\linewidth]{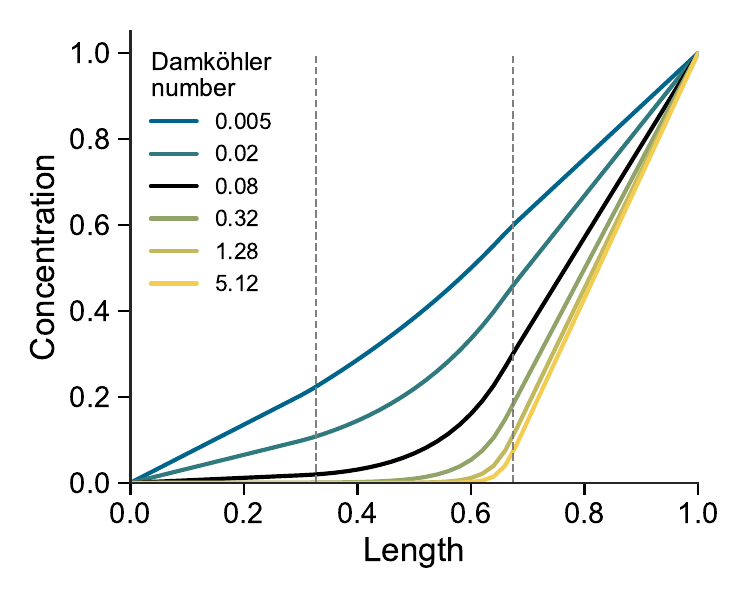}
    \caption{Simulation results of semi-permeable diffusion with a reactive membrane in between the dashed lines. The impact of variations of the reaction rate is shown. The black line is the common baseline. The composite fraction $\eta^\mathrm{RBC}=0.1$.} 
    \label{fig:reactiveMembraneDa}
\end{figure}

\begin{figure}[t]
    \centering
    \includegraphics[width=0.7\linewidth]{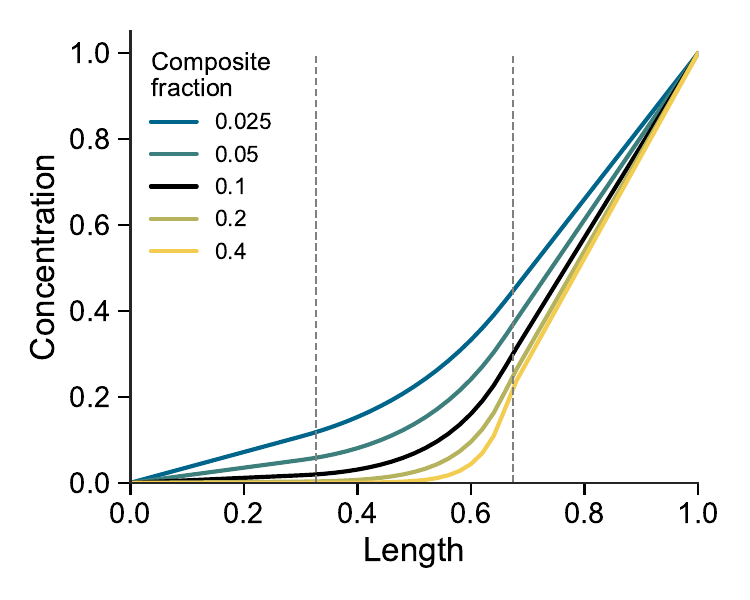}
    \caption{Simulation results of semi-permeable diffusion with a reactive membrane in between the dashed lines. The impact of variations of the composite fractions $\eta^\mathrm{RBC}$ are shown. The black line is the common baseline. The reaction rate is constant at $\mathrm{Da}=0.08$.} 
    \label{fig:reactiveMembraneCF}
\end{figure}

In this rewritten form, the PBBF clearly shows the advection-diffusion through porous media, with $\Omega^\mathrm{PBB}\left[ f_i \right] = \Omega^\mathrm{BGK}\left[f_i^\mathrm{BGK}\right] + \Omega^\mathrm{BB}\left[f_i^\mathrm{RBC}\right]$, and has a source term of $+ 2\eta^\mathrm{RBC} \eta^\mathrm{ABB} \Omega^\mathrm{ES}\left[ f_{\bar{i}} \right]$. The $\Omega^\mathrm{PBB}\left[ f_i \right]$ simulates diffusion with an effectively lower diffusivity due to tortuosity \cite{walshNewPartialbouncebackLatticeBoltzmann2009}. The $\eta^\mathrm{RBC}$ already contains $\tau$ dependent corrective terms.
This new collision step is showcased here as a reactive membrane in a quasi-1D reaction-diffusion problem. Simulations are performed on a 2D domain of size $10\times50$, periodic in the y-axis, with two Dirichlet boundaries defining the concentration to $\rho=0$ at $x=0$ and $\rho=1$ at $x=50$. A semi-permeable membrane is initialized from $x=16$ to $x=33$. The equilibrium concentration is $\rho^\mathrm{eq}=0$, it is defined that $\bci\cdot\mathbf{n}=1$ for all $i$, the relaxation time is $\tau=0.8$. A variation in the Damköhler number and composite fraction $\eta^\mathrm{RBC}$ is performed. The resulting concentration profile are shown in \cref{fig:reactiveMembraneDa,fig:reactiveMembraneCF}.

From the concentration profiles seen in \cref{fig:reactiveMembraneDa,fig:reactiveMembraneCF}, there are three distinct regions: the membrane indicated by the dashed box, and the two freely diffusive regions to the left and right of the membrane. In the regions to the left and the right of the membrane, the concentration profile is linear, as one would expect for a constant diffusion coefficient. Inside the membrane, the concentration profile is increasing exponentially.

Both the Damköhler number and the composite fraction impact the reactivity of the membrane and, thus, the resulting concentration profile. The exponential is strongly determined by the composite fraction $\eta^\mathrm{RBC}$ (cf.~\cref{fig:reactiveMembraneCF}). The Damköhler number has a larger impact on the concentration decrease than the composite fraction. 

Since for advection-diffusion, the bounceback composite fraction is related to the tortuosity of the medium \cite{walshNewPartialbouncebackLatticeBoltzmann2009}. Changing this fraction allows studying the impact of a change in the tortuosity. 
%This can be applied to study various systems, e.g.\ particles flowing through filters, precipitation in nano-porous battery electrodes, or heat flow through a radiator. 

\section{Conclusion}
We proposed a framework for composite collisions within the lattice Boltzmann method. This framework decomposes collision steps into component collision rules. For linear collision rules, this can be interpreted as splitting the total populations into component populations, applying a specific collision operator to each, before finally recombining them. This enables modeling of more complex physical phenomena through combination of various simple collision operators. Special care has to be taken for the inclusion of forces in the scheme and we provide a general recipe for this aspect.

Benefits of the approach are demonstrated on several examples.
It is shown how using this \MN\ approach may serve as a valuable analysis method. It is shown that the Robin boundary condition is a composition of bounceback and anti-bounceback collision rules. This insight allows an easier extension, to include moving walls or more accurate collision rules. 

Additionally, the \MN\ was used to create two novel composite collision steps: the partial Robin boundary conditions (PRBC) and partial bounceback with fluxes (PBBF). This demonstrates how the \MN\ can be used to synthesize new collision steps to capture non-trivial physical phenomena. 

Several research avenues for the \MN\ remain open. One possibility is to account for inter-collision interaction. Maybe it is possible to have microscopic velocity dependent composite fractions i.e.\ $\eta_i^\colInd$. And finally, investigating the \MN's applicability and limits for non-local collision rules. Overall, the framework provides a systematic approach to derive lattice Boltzmann methods for complex physical processes. 

\begin{acknowledgments}
\vspace{1em}\noindent The authors gratefully acknowledge financial support by the Federal Ministry of Education and Research (BMBF) within the project ``SulForFlight'' under the Grant No.03XP0491A. M.L. and B.K. gratefully acknowledge financial support from the European Union's Horizon 2020 Research and Innovation Programme within the project ``DEFACTO'' under the grant number 875247. This work contributes to the research performed at CELEST (Center for Electrochemical Energy Storage Ulm-Karlsruhe).
\end{acknowledgments}

\section*{Author Contributions}
\noindent\textbf{Julius Weinmiller:} 
Conceptualization;
Investigation;
Methodology;
Software;
Visualization;
Writing – original draft; \\
\textbf{Benjamin Kellers:} 
Conceptualization;
Methodology;
Validation;
Writing – review \& editing;\\
\textbf{Martin P. Lautenschlaeger:}
Conceptualization;
Writing – review \& editing; \\
\textbf{Timo Danner:} 
Conceptualization;
Funding acquisition;
Project administration;
Supervision;
Writing – review \& editing; \\
\textbf{Arnulf Latz:}
Funding acquisition;
Supervision;
Writing – review \& editing;

\appendix
\renewcommand\thefigure{\thesection.\arabic{figure}}

\section{Comparing RBC and PBB composite fraction}
\label{app:compareCompositeBB}

In the literature on partial bounceback methods, the method of Walsh, Burwinkle and Saar \cite{walshNewPartialbouncebackLatticeBoltzmann2009} also used the fullway bounceback collision step. They found that their composite fraction ($n_s$ in their work) is
\begin{equation}
    \label{eq:eta_Walsh}
    \eta^\mathrm{BB}=\frac{\tau-\Dt/2}{2k\Dt/\cs^2 +  (\tau-\Dt/2) }=\frac{1}{2k\Dt/\nu + 1}, 
\end{equation}
where $k$ is the permeability of the unresolved solid/fluid mixture.

The bounceback composite fraction of RBC (cf.\ \cref{eq:RBC_eta}) is reiterated here
\begin{equation}
    \eta^\mathrm{BB} = \frac{1}{\tau \cdotnConst k_\mathrm{r}/ D + 1 }. \nonumber
\end{equation}
For a fluid, the viscosity is the momentum diffusivity. Hence it makes sense that instead of viscosity the scalar diffusivity appears here. The transfer rate $k_\mathrm{r}$ is then analogous to the permeability $k$. 
%The similarity of the $\eta$ definitions indicates a general theme.

\section{PRBC paramter $\prbcFit$ sensitivity}
\label{app:PRBCsensitivity}
\setcounter{figure}{0}

\begin{figure}[t]
    \centering
    \includegraphics[width=0.7\linewidth]{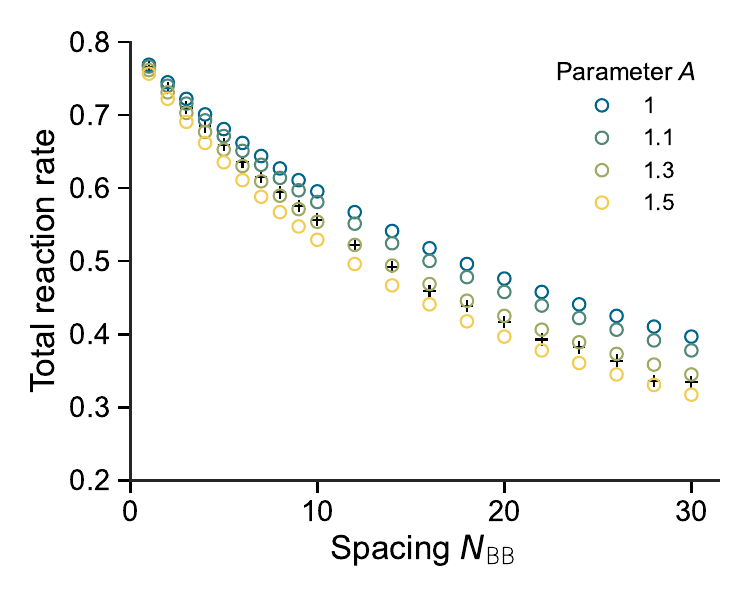}
    \caption{Sensitivity of the fitting parameter $\prbcFit$ on simulated non-dimensional reaction rated. The reaction rates of the boundary condition using the alternating BC (black pluses) vs PRBC (colored circles) are shown. The spacing $N_\mathrm{BB}$ is varied between 1 and 30. The Damköhler number $\mathrm{Da}=50$ was simulated with fitting parameters $\prbcFit=\{1,\ 1.1,\ 1.3,\ 1.5\}$.}
    \label{fig:PRBCsensitivity}
\end{figure}

In the PRBC, the composite fraction $\eta^\mathrm{RBC}$ is dependent on a fitting parameter $\prbcFit$ (cf.~\cref{eq:etaPRBC}). Here, the sensitivity of this parameter is shown for one simulation. The $\mathrm{Da}=50$ simulation is used, since it is neither diffusion, nor reaction limited, and thus should be most sensitive to variation of $\prbcFit$. The results are shown in \cref{fig:PRBCsensitivity}. 

%\todo{Update fig if fit name changes}

From \cref{fig:PRBCsensitivity}, it can be concluded that the fitting parameter has the largest impact when the spacing is very large. For small spacing, the impact is negligible. Even for the worst case with large spacing, the simulation is more sensitive to a decrease in the reaction rate, than to a variation in the fitting parameter.

\bibliography{ZoteroLib}

\end{document}